\documentclass[preprint]{aastex}

\shorttitle{Globular cluster colors in NGC 3311: normal} 
\shortauthors{Brodie, Larsen \& Kissler-Patig} 

\begin{document}

\title{ A new look at globular cluster colors in NGC 3311 and the case
for exclusively metal--rich globular cluster systems}

\author{J.P. Brodie and S.S. Larsen}
\affil{UC Observatories / Lick Observatory, University of California, Santa 
Cruz, CA 95064, USA}
\email{brodie@ucolick.org, soeren@ucolick.org}

\and 

\author{M. Kissler-Patig} 
\affil{European Southern Observatory, Karl-Schwarzschild-Str.~2, 85478 Garching
b. M\"unchen, Germany}
\email{mkissler@eso.org}

\begin{abstract}
NGC 3311, the central cD galaxy in the Hydra cluster, was previously
thought to host the most metal--rich globular cluster system
known. Ground--based Washington photometry had indicated the almost complete
absence of the population of globular clusters near [Fe/H] $\sim -1$ dex,
normally dominant in the metallicity distribution functions of giant
elliptical galaxies.  Lacking the normal metal--poor globular cluster
population, NGC 3311 was an outstanding exception among galaxies, not
easily understood under any of the current globular cluster formation
scenarios. Our HST/WFPC2 data yield normal globular cluster colors and
hence metallicities for this galaxy.  We find a bi--modal color distribution
with peaks at $(V-I)_o=0.91\pm0.03$ and $1.09\pm0.03$, corresponding to
[Fe/H]$\sim -1.5$ and $-0.75$ dex (somewhat dependent on the choice of the
conversion relation between color and metallicity). We review the evidence
for exclusively metal--rich globular cluster systems in other galaxies and
briefly discuss the implications for our understanding of globular cluster
and galaxy formation.

\end{abstract}



\section{Introduction}

Globular cluster systems are thought to be 
good tracers of the star--formation history of
their host galaxies (see recent reviews by Ashman \& Zepf 1998, Harris
2000, Kissler-Patig 2000). A variety of theories exist to explain
the presence of distinct globular
cluster sub--populations in many galaxies.
For example, the bi--modal
color distributions seen in many bright early--type galaxies can
result from spiral--spiral mergers (Ashman \& Zepf 1992), in--situ
multi--phase galaxy/globular cluster formation (Forbes, Brodie \&
Grillmair 1997; Kissler-Patig, Forbes \& Minniti 1998a) and/or 
accretion/stripping mechanisms (C\^ot\'e, Marzke \& West 1998; Hilker, Infante \& Richtler 1999).
Implicit in these theories 
is the presence of an old, {\it metal--poor} (halo?) globular
cluster population which should exist in essentially all galaxies. 

cD galaxies have ``abnormally'' high
specific frequencies (number of globular clusters per unit galaxy light,
Harris \& van den Bergh 1981) and, generally, these galaxies have
similar (large) numbers of blue and red globular clusters or a preponderance
of blue clusters (e.g.~Forbes, Brodie \& 
Grillmair 1997). Their high specific frequencies may have resulted from
augmentation of the {\it low--metallicity} component by galaxy accretion
or globular cluster stripping. Alternatively, star 
formation may have been truncated at an
early stage, i.e.~they do not have too many globular clusters but rather
too few stars \citep{bla97,har98,mac99}.

The globular cluster system of NGC 3311 is of particular interest for
understanding galaxy and globular cluster formation because it is the next
nearest cluster central cD after M87 in Virgo and NGC 1399 in Fornax and
its specific frequency is one of the highest known \citep{har83,mac95}.

In this context the result of \citet{sec95} was surprising. Using
Washington (C-T$_1$)$_o$
photometry from the CTIO 4m, these authors found that NGC 3311, the central cD
galaxy in the Hydra cluster, had an unusually red globular cluster system
with a median metallicity of [Fe/H]=$-$0.31 dex. 
A third of the clusters were estimated to
exceed solar abundance and the population of clusters near [Fe/H]$\sim -1$
dex, normally dominant in the metallicity distribution functions (MDFs) of
giant elliptical galaxies (see above reviews), was found to be almost totally 
absent in NGC 3311. The MDF was found to be bi--modal at the 97.6\% confidence 
level in a KMM analysis, with peaks at [Fe/H]=$-0.5$ and +0.15.  These results 
meant that NGC 3311 was host to the most metal--rich globular cluster system 
known and required extensive pre--enrichment of the entire halo before 
essentially any of the stars or clusters formed.
This posed serious problems for all three of the main contending
scenarios for the formation of globular cluster systems. 

Here we examine the color distribution of the NGC 3311 globular cluster
system using new data from the HST WFPC2 camera and show that this distribution
is, in fact, quite normal and that the system is not especially metal--rich.
A paper examining other properties of the NGC 3311 globular cluster system
is in preparation and will include comparisons with a number of other
galaxies from our HST database.

\section{Data and Reduction}

  Our data consist of Cycle 6 WFPC2 observations.
The integration times were 3700\ s and 3800\ s in the F555W and F814W 
filters, respectively, split into 4 exposures for each filter. The 
raw exposures were processed by the standard STScI pipeline and subsequent
reduction and analysis were performed using IRAF\footnote{IRAF is distributed
by the National Optical Astronomical Observatories, which are operated by
the Association of Universities for Research in Astronomy, Inc.~under contract
with the National Science Foundation}. 

  The individual exposures were aligned using the {\tt imshift} task and
were then combined using {\tt imcombine}.  Cosmic Ray events were
eliminated by setting the {\tt reject} parameter in {\tt
imcombine} to {\tt crreject}.

  A first set of background--subtracted images was generated by smoothing
the original images with a $15\times15$ box median filter and 
subtracting the smoothed images from the original images. Objects were 
detected on the background--subtracted images
using the task {\tt daofind} in the {\tt DAOPHOT} package within IRAF 
\citep{ste87} and were then removed from the 
original images using the {\tt ishape} algorithm \citep{lar99}. The 
object--subtracted images were then median--filtered once again and subtracted 
from the original images, providing our final set of background--subtracted 
images for further analysis.

  A new object--detection process was performed on the combined,
background--subtracted images in both F555W and F814W.
The two object lists were then matched and aperture photometry was obtained
using the {\tt phot} task in DAOPHOT, using an $r=2$ pixels aperture. At
the distance of NGC~3311 (distance modulus $\sim 33$, assuming pure
Hubble flow, $V_0\sim3400$ km s$^{-1}$ (NASA/IPAC Extragalactic Database), and
$H_0=75$ km s$^{-1}$ Mpc$^{-1}$) one WF camera pixel corresponds to a
linear scale of $\sim 19$ pc, so globular clusters are unresolved on our
HST images.  The use of a small aperture to minimize random errors is thus
unlikely to introduce any {\it systematic} errors due to extendedness of
the objects. We considered only the
objects in the WF chips. Since our WFPC2 pointings were
centered on the nucleus of NGC~3311 we thus avoid the central region of the
galaxy out to a radius of $\sim 3.5$ kpc. 
The PC frames contain $\sim$20\% of the total number of objects in
the 4 WFPC2 chips. Since we have so many clusters
from the WF chips, including the PC objects does not significantly
improve the statistics or alter the
location of the $V-I$ peaks. PC objects do, however, require different
aperture corrections and they are excluded to avoid any possible
resulting systematic errors.

Figure 1 shows NGC 3311, the nearby galaxy NGC 3309, and the location of
the Wide Field camera chips for this pointing.

  The photometry was calibrated to standard $V$ and $I$ magnitudes following 
\citet{hol95}.  We calculated the aperture corrections 
between our $r=2$ and the standard Holtzman et al.~$r=5$ pixels apertures by 
measuring objects brighter than $V=25$ in both apertures. We found a small 
difference in the F555W and F814W aperture corrections of 
$\Delta(V-I)_{2\rightarrow5} = 0.026$ magnitudes. Although the
{\it differences} between the aperture corrections in the two filters are in
good agreement with those found by other authors \citep{puz99,whi97}, the aperture 
corrections themselves are about 0.1 magnitude larger than the values obtained
by Whitmore et al. This is most likely due to the resampling of the PSF that 
is done by {\tt imshift} when aligning the images.  As a check, we also 
carried out photometry on a combined subset of 2 images in each filter where 
no shifts were required; in this case we found aperture corrections 
in very good agreement with those of Whitmore et al. 
However, we are not going to 
refer to absolute magnitudes in this letter but only to colors, and we
will use the photometry based on 4 combined images because of the better
noise characteristics and superior CR rejection. 

We checked our results by performing an ALLSTAR PSF--fitting reduction of
the data.  There is an insignificant affect on the location of the $V-I$
peaks (at the level of 0.01 mag., well within our estimated uncertainties)
and no subsequent conclusions are affected by choosing aperture rather than
PSF photometry.

\section{Revised colors and metallicities for globular clusters in NGC 3311}

\subsection{Colors}

Figure 2, upper panel, is the reddening--corrected color--magnitude diagram
of all objects in the Wide Field camera chips. An extinction correction of
E$_{V-I}$=0.13 (from E$_{B-V}$=0.079, A$_V$=0.24, Schlegel et al.~1998) has
been applied.  Note that \citet{sec95} used E$_{B-V}$=0.045 from
\citet{bur84}, corresponding to E$_{V-I}$=0.072, i.e.~there is a systematic
difference of 0.05 mag.~in E$_{V-I}$ between our colors and the
corresponding colors quoted by Secker et al. Although it is in the right
sense, this difference is not nearly sufficient to account for the
disagreement.

Globular cluster candidates were selected from within the range 0.6 $< V-I
<$ 1.4. There are 1971 candidates with $V<26$ and 882 with $V<25$. A KMM
test \citep{ash94} finds peaks at $V-I = 0.91$ and $1.09$ for the magnitude
interval $22<V<25$ (corresponding roughly to $-11<M_V<-8$) and the
distribution is bi--modal at the 94.4\% confidence level.  Peaks are found
at $V-I$ = 0.87 and 1.09 for the interval $22<V<26$ with bi--modality
indicated at the $>$99.9\% confidence level. Changing the magnitudes thus
yields consistent values for the $V-I$ peaks and we estimate that the peak
colors are accurate to about 0.03 magnitudes.  About half of the clusters
belong to the blue population and half to the red. The lower panel in
figure 2 shows a color histogram for the globular clusters down to a
magnitude limit of V=25.  Overplotted are the two best--fitting Gaussians
from the KMM analysis as well as their sum.

DIP statistics \citep{geb99} indicate the probability that the distribution
is not unimodal. For the magnitude interval $22<V<24$ the DIP value is
99.1\%, for $22<V<25$ it is 98.7\% and for $22<V<26$ it is 89.1\%. The 50\%
completeness limit for the data is at $\sim$26.5 mag.~in V, but the
observational scatter clearly increases substantially below V=25.

\subsection{Metallicities}

To convert our $V-I$ colors to metallicities we use the relations derived
by \citet{cou91} and \citet{kis98b}. These relations
provide the best calibrations for the low and the high metallicity
ranges, respectively.  Using the $V-I$ values for our brighter magnitude
interval, i.e.~$0.91$ and $1.09$, the conversion relations yield 
metallicities of [Fe/H]$\sim -1.50$ and $-0.59$ dex and [Fe/H]$\sim -1.52$ 
and $-0.94$ dex, for the blue and red peaks respectively. The errors
introduced by the internal uncertainties of the conversion relations are 
around 0.3 dex. Note that if we used the reddening
value adopted by \citet{sec95} the peak colors would redden to
$(V-I) \sim 0.96$ and 1.14, corresponding to metallicity
shifts of only $\sim +0.2$ dex.  These
values we derive are typical for bright elliptical galaxies (Forbes, Brodie \& Grillmair
1997). In particular, the peaks in M87 are found at $V-I$=0.95 and 1.20
mag.  \citep{kun99} corresponding to [Fe/H]$\sim -1.4$ and $-0.6$
dex (using the Kissler-Patig et al.~relation).  The peaks in NGC 1399 occur
at $V-I=$0.99 and 1.18 \citep{kis98b}, corresponding to
[Fe/H]$\sim -1.3$ and $-0.6$ dex. If anything, the metal--rich peak in NGC 3311
may have somewhat lower metallicity than the metal--rich peaks in the central
cDs of the Virgo and Fornax clusters.

The data of \citet{sec95} were obtained under non--photometric conditions.
As described in their text, these authors clearly made every attempt to
correctly calibrate their frames using conventional and appropriate
techniques. However, given that our HST colors agree well with those for
other central cD galaxies' globular cluster systems, it seems likely that
the Secker et al.~ results suffer from a zero--point error.

\section{Discussion}

\subsection{Do systems exist with no metal--poor globular clusters?}

\citet{zep95} studied the globular clusters in NGC 3923 and reported an
exceptionally red color distribution for this galaxy's luminosity. We note,
however, that these observations were made on the same CTIO 4m,
non--photometric, observing run as NGC 3311 and were calibrated {\it a
posteriori} with the same data. We suggest that these observations be
revisited in light of the differences found for NGC 3311.

In the Coma cluster, \citet{woo00} found, from HST data, that
IC 4051 has a high mean globular cluster metallicity ([Fe/H]$\sim -0.3$
dex). The color distribution appears to be unimodal with a narrow dispersion, 
although Woodworth and Harris were not able to exclude the possibility
of a bi--modal distribution with two close
peaks and a metal--poor population around [Fe/H]$\sim -1.0$ dex.
The specific frequency of IC 4051 is high 
(S$_N$=12$\pm$3), comparable to those of cluster central cDs, although
IC 4051 is neither the central galaxy in this rich galaxy cluster nor is
it exceptionally luminous (M$_V -21.9$).

In their survey of archival HST data, \citet{geb99}
found evidence for systems dominated by metal--rich globular clusters.
However, none of these cases were as extreme as NGC 3311 was previously
thought to be. The
existence of a blue (metal--poor) population in these galaxies is not
excluded by these data (because of small sample statistics) but Gebhardt \&
Kissler-Patig suggest that it is not likely to be significant, as in
IC 4051.

In summary, no system has yet been observed to have a globular cluster
color distribution as extremely red as the previous claim for NGC 3311.  In
particular, {\it no galaxy has yet been convincingly shown to be entirely
lacking globular clusters with metallicities below [Fe/H]$\sim -1.0$ dex}.

\subsection{Implications for formation models of globular cluster
systems} 

NGC 3311 represented the best case of an almost exclusively metal--rich
globular cluster population.  As such it raised serious questions about our
understanding of globular cluster and galaxy formation, since the existence
of a metal--poor population is implicit in all the current scenarios. 

In the merger picture an elliptical galaxy is formed from the merger of two
gas--rich spiral galaxies. The resultant galaxy contains a blue,
metal--poor population of globular clusters from the progenitor spirals and
a metal--rich population formed during the merger. In this scenario the
metal--poor population should peak at the median metallicity of the
globular cluster systems of spiral galaxies, i.e.~[Fe/H] $\sim -1.5$
dex. It is generally recognized, though, that mergers are unlikely, by
themselves, to result in high specific frequency galaxies, such as NGC 3311
(Ashman \& Zepf 1998; Forbes, Brodie \& Grillmair 1997). In the
multi--phase picture, the blue population is formed in a pre--galaxy
phase, or during galaxy assembly, from relatively unenriched gas and so, by
definition, is metal--poor. The red globular cluster population and the
bulk of the galaxy stars are formed later from enriched material.  In the
accretion model, the original elliptical galaxy (formed in a single burst)
accretes smaller (lower metallicity) galaxies with their retinues of
low--metallicity globular clusters.  Giant galaxies at the centers of rich
galaxy clusters may also be expected to strip metal--poor globular clusters
from the outskirts of neighboring galaxies. Because of the globular cluster
mean metallicity -- parent galaxy luminosity relation \citep{bro91,for96}
accreted/stripped globulars must be, on average, of lower metallicity than
those belonging to the ``seed'' elliptical.  \citet{cot98} show how
accretion processes and the luminosity function of galaxies will lead to
bi--modal globular cluster systems in bright galaxies with blue and red
peaks at [Fe/H] $\sim -1.5$ and $\sim -0.5$ dex, as generally observed.

Sketchy arguments can be imagined to accommodate an absence of
metal--poor clusters within any of the above formation scenarios.
To produce only, or predominantly, metal--rich clusters under the ``in situ''
scenario would require rapid star formation prior to cluster formation in
the same star formation event. In a deep gravitational potential,
the metals produced by the stars are retained and the average metallicity
of the system is driven to a  high value (see also Woodworth \& Harris 2000),
conceivably before any clusters are formed.
In accretion scenarios, the
metal--poor population in the final galaxy would be minimized if small
fragments/dwarf galaxies were absent from the galaxy's environment.
Note, though, that such a situation is less likely for a central cD galaxy.
Late mergers are expected to provide a significant population of
metal--poor globular clusters originating in the progenitor galaxies,
predominantly metal--rich systems could be explained if the progenitors
were gas--rich but cluster--poor (see also Gebhardt \& Kissler-Patig
1999). Stripping can result in the preferential removal of metal--poor
clusters if the metal--poor population is more extended than the
metal--rich one.  Since stripping will act more efficiently on the more
extended population, it could result in systems that are biased towards
high metallicities.  In this context, it is interesting to note that the
halo light of IC 4051 appears truncated beyond $\sim$ 30 kpc \citep{jor92}.

Preferentially metal--rich globular cluster systems can be accommodated by
adjusting the existing scenarios.  An exclusively metal--rich globular
cluster population, should one be discovered, would probably be most easily
accommodated in a single collapse model -- an ``in situ'' galaxy formation
scenario with rapid star--formation preceding cluster formation.
  
\section{Summary}

We have shown that the color 
distribution of globular clusters in NGC 3311
is normal for bright elliptical galaxies. It is bi--modal with peaks at
$V-I\sim0.91$ and $1.09$, corresponding to metallicity peaks at around
[Fe/H$\sim -1.5$ and $-0.75$ (the precise values being dependent on the
choice of the conversion relation between color and metallicity). This
range of metallicities is normal for bright elliptical galaxies, a
result which contradicts an earlier claim that NGC 3311 might host an extremely
metal--rich globular cluster system.

We suggest that it is worth revisiting the globular cluster system of NGC
3923 whose globular cluster system was reported to be very red on the basis
of observations made on the same run which produced the NGC 3311 results.

Although the evidence for {\it exclusively} metal--rich globular cluster
systems has been weakened, there are still cases of galaxies whose globular
cluster systems may lack a significant metal--poor component.  We have
briefly discussed the implications of such systems for our understanding of
globular cluster and galaxy formation and have concluded that, with some
adjustments, they can be explained under existing scenarios.

\acknowledgments

We thank John Huchra for his help and useful suggestions and Duncan Forbes,
Carl Grillmair and Ken Freeman for their contributions.  This work was
supported by HST grant GO.06554.01-95A, National Science Foundation grant
number AST9900732 and Faculty Research funds from the University of
California, Santa Cruz.

\clearpage

\plotone{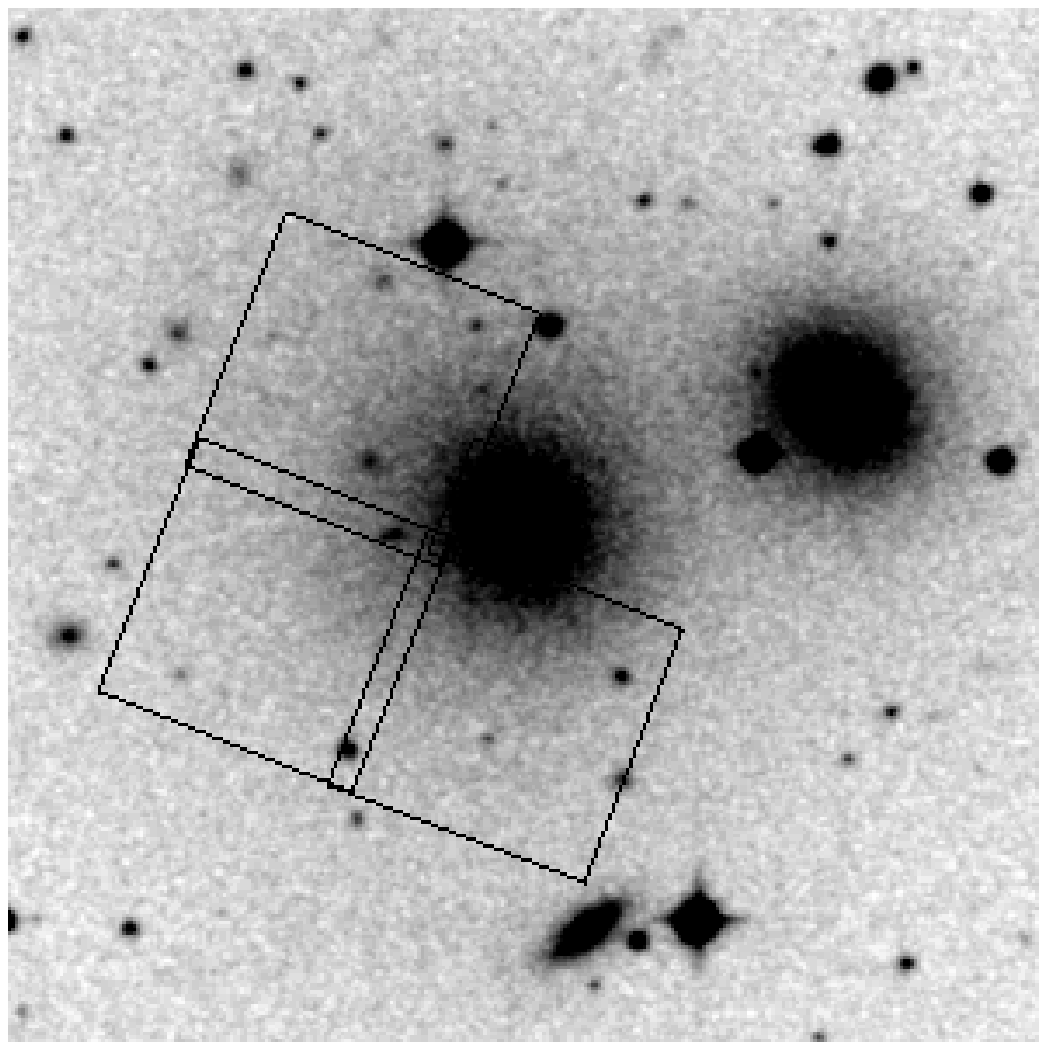}
\figcaption[brodie.fig1.ps]{NGC 3311 and the nearby galaxy, NGC 3309, with 
the location of the WF chips superimposed. \label{fig1}}

\plotone{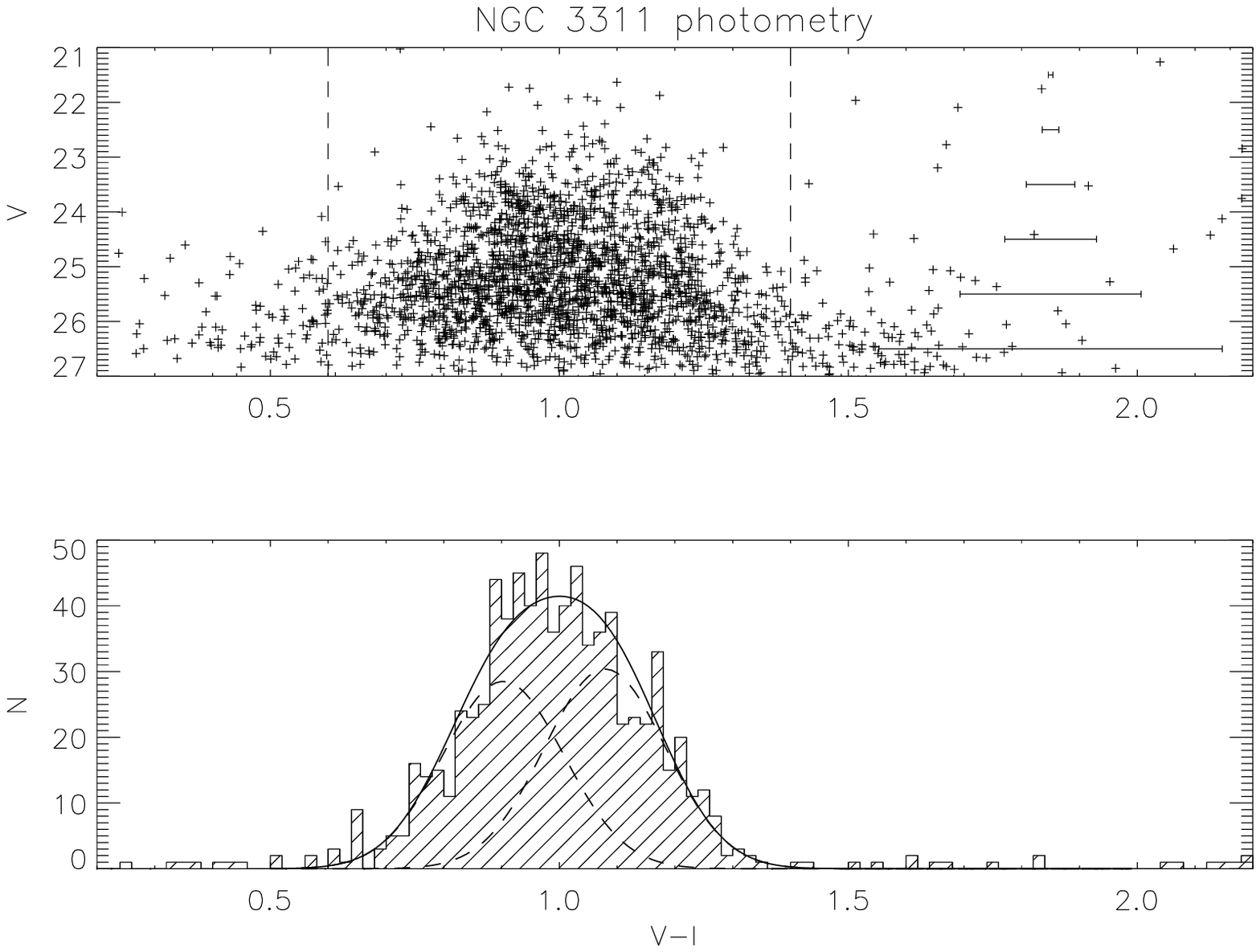}
\figcaption[brodie.fig2.ps]{Upper panel: $V-I,V$ color--magnitude diagram for objects in 
NGC~3311. The photometry has been dereddened using $A_V = 0.24$ 
\citep{sch98}. The photometric errors are indicated by the
error bars at $V-I$=1.85. Lower panel: $V-I$ histogram for objects in 
NGC~3311 brighter than V=25. Overplotted are the two best--fitting
Gaussians from the KMM analysis along with their sum. \label{fig2}}

\end{document}